\input{aipcheck}
\documentclass[
    ,final            
  ]
  {aipproc}

\layoutstyle{6x9}
\usepackage{graphicx}
\usepackage{rotating}
\usepackage{geometry}
\usepackage{graphics}

\def\ba{\begin{eqnarray}}
\def\ea{\end{eqnarray}}
\def\ba{\begin{eqnarray}}
\def\ea{\end{eqnarray}}
\def\atp{\frac{\alpha_s(Q^2)}{2\pi}}
\begin{document}

\title{NNLO Logarithmic Expansions and High Precision Determinations of the
QCD background at the LHC: The case of the Z resonance}


\author{$^1$Alessandro Cafarella, $^{2,3} $Claudio Corian\`o and $^2$Marco Guzzi}{
  address={ $^1$Institute of Nuclear Physics, NCSR ``Demokritos'',15310 Athens, Greece \\
  $^2$Dipartimento di Fisica, Universit\`a del Salento and INFN Sezione di Lecce, \\ via Arnesano 73100, Lecce, Italy\\
  $^3$Department of Physics and Institute of Plasma Physics, University of Crete, Heraklion, Greece.}
}


\begin{abstract}
New methods of solutions of the DGLAP equation and their implementation through NNLO
in QCD are briefly reviewed. We organize the perturbative expansion that describes in $x$-space the evolved parton distributions in terms of scale invariant functions, which are determined recursively, and logarithms of the ratio of the running couplings at the initial and final evolution scales. Resummed solutions are constructed within the same approach and
involve logarithms of more complex functions, which are given in the non-singlet case. Differences in the evolution schemes are shown to be numerically sizeable and intrinsic to perturbation theory. We illustrate these points in the case of
Drell-Yan lepton pair production near the Z resonance, analysis that can be extended to searches of extra $Z^{\prime}$. We show that the reduction of the NNLO cross section compared to the NLO prediction may be attributed to the NNLO evolution.
 \footnote{Presented by M. Guzzi at {\it QCD@work 2007}, Martina Franca, Italy, 16-20 June 2007.}

\end{abstract}

\maketitle

\section{Introduction}

QCD will enter its precision era with the advent of the LHC.
For this reason, from the theory side, precise determinations of some key observables at LHC energies
will be essential both for the discovery of new physics and for  QCD partonometry.
One of the first searches that we expect to be performed at the LHC will be the study of the rapidity and the invariant mass
distributions in Drell-Yan on the Z resonance and at larger invariant mass values,  which will allow to calibrate the parton distributions
functions (pdf's) at the new (higher) factorization scales available at the new collider.
In return, these first studies of the background will be crucial in order to quantify the luminosity of the machine and for the investigation of possible
extensions of the Standard Model, from the ``simplest'' ones concerning the Higgs or extra neutral gauge interactions, to more complex ones involving supersymmetry or popular scenarios such as the those concerning theories with {\it extra dimensions}.
Given the high accuracy of the current computations, which in some case have reached  the few-percent level of precision, and the rather small variations
observed in moving from next-to-leading order (NLO) to next-to-next-to-leading order (NNLO) in some key observables,
it is rather natural to ask whether the sistematic errors due to the implementation of the evolution of the pdf's are of the same order of the new perturbative contributions coming from the hard scatterings.
We answer these points by presenting a new method of solutions of the DGLAP equation that we have developed
and tested  in several previous works that we call {\em the method of the logarithmic expansions}.  
The $x$-space solutions are classified in terms of the different types of accuracies selected in 
the construction of the evolved pdf's, solutions that bring in a systematic uncertainty in the 
prediction of the hadronic observables. The theory behind this new approach can be found in \cite{CCG1},\cite{CCG2}.

\section{The NNLO Evolution}

Studies of  hard scattering cross sections such as the invariant mass distribution of the lepton pair in
Drell-Yan $d\sigmaƒ/dM$ \cite{Van_Neerven1}, and the rapidity distribution $d\sigmaƒ/dM dY$  \cite{Anastasiou}
are available and can be used as tests for the NNLO evolution. Some of these computation have been performed before that the analytical NNLO DGLAP were computed (see \cite{vogt1}).

We recall that the perturbative expansion of the DGLAP splitting
functions and of the $\beta$-function take the generic form (to the m-th order)
\ba
\label{kernel_NmLO}
&&{P^{N^m LO}(x,Q^2)}=\sum_{i=0}^{m}\left(\atp\right)^{i+1} P^{(i)}(x,Q^2),
\nonumber\\
&&\frac{\partial \beta^{N^m LO}(\alpha_s(Q^2))}{\partial \log{Q^2}}=
\sum_{k=0}^{m}\left(\frac{\alpha_s(Q^2)}{4\pi}\right)^{k+2}\beta_{k},
\ea
with $\beta_k$ being the corresponding coefficients of the $\beta$ function which have been summarized in
\cite{CCG1}. Leading order (LO), NLO and NNLO expansions correspond to the cases $m=0,1,2$ respectively.

We solve the DGLAP equation by formulating a general ansatz in (Bjorken) $x$-space which holds to any
perturbative order and which allows us to have a control on the accuracy in terms of
powers of $\alpha_s$ the strong coupling constant. It is given by \cite{CCG1}
\begin{equation}
f_{N^{\kappa' LO}}(x,\alpha_s)\big|_{O(\alpha_s^{\kappa'})}=
\sum_{n=0}^\infty\left(A^0_n(x) + \alpha_s A^1_n(x)+ \alpha_s^2 A^2_n(x) + \dots +
\alpha_s^{\kappa'} A^{\kappa'}_n(x)\right)
\left[\ln{\left(\frac{\alpha_s(Q^2)}{\alpha_s(Q^2_0)}\right)}\right]^{n}.
\nonumber\\
\label{truncatedseries}
\end{equation}
The expansion is valid 
both in the non-singlet and in the singlet sectors, and the scale-invariant coefficients $A^k_n(x)$  are determined by solving
a chain of recursion relations obtained by substituting the ansatz into the DGLAP equation. We have implemented the method in a numerical code \textsc{Candia}, which will be made available soon.

\subsection{The logarithmic resummation of the DGLAP solutions}
When we move from logarithmic expansions of the form (\ref{truncatedseries}) to exact solutions, we are actually performing a resummation of the perturbative expansion of the pdf's. This point has been discussed in our original works 
\cite{CCG1,CCG2}, to which we refer for more details. Here we just quote the result.

The general solution can be written in terms of $A^{\prime}_n, B^{\prime}_n, C^{\prime}_n $,
coefficients that will be calculated by a chain of recursion
relations \cite{CCG1} giving
\begin{eqnarray}
f(x,Q^{2}) & = & \left(\sum_{n=0}^{\infty}\frac{A'_{n}(x)}{n!}\mathcal{L}^{n}\right)_\otimes
\left(\sum_{m=0}^{\infty}\frac{B'_{m}(x)}{m!}\mathcal{M}^{m}\right)_\otimes
\left(\sum_{p=0}^{\infty}\frac{C'_{p}(x)}{p!}\mathcal{Q}^{p}\right)_\otimes f(x,Q_0^2)\nonumber \\
& = & \sum_{s=0}^{\infty}\sum_{t=0}^{s}\sum_{n=0}^{t}\frac{A'_{n}(x)\otimes
B'_{t-n}(x)\otimes {C'_{s-t}(x)}}{n!(t-n)!(s-t)!}
\otimes f(x,Q_0^2)\,\mathcal{L}^{n}\mathcal{M}^{t-n}\mathcal{Q}^{s-t}\nonumber \\
& = & \sum_{s=0}^{\infty}\sum_{t=0}^{s}\sum_{n=0}^{t}
\frac{ D_{t,n}^{s}(x)}{n!(t-n)!(s-t)!}\mathcal{L}^{n}\mathcal{M}^{t-n}\mathcal{Q}^{s-t},
\label{eq:NNLOansatz}
\end{eqnarray}
where $\otimes$ is the standard convolution product of the DGLAP
and where
\begin{equation}
D_{t,n}^{s}(x)= A'_{n}(x)\otimes
B'_{t-n}(x)\otimes C'_{s-t}(x)\otimes f(x,Q_0^2).
\end{equation}
The explicit expression of the functions appearing in the ans\"atz are
\begin{eqnarray}
\mathcal{L} & = & \log\frac{\alpha_s}{\alpha_{0}},\\
\mathcal{M} & = & \log\frac{16\pi^{2}\beta_{0}+4\pi\alpha_s\beta_{1}
+\alpha_s^{2}\beta_{2}}{16\pi^{2}\beta_{0}+4\pi\alpha_{0}\beta_{1}+\alpha_{0}^{2}\beta_{2}},\\
\mathcal{Q} & = & \frac{1}{\sqrt{4\beta_{0}\beta_{2}-\beta_{1}^{2}}}\arctan \chi,\\
\end{eqnarray}
and 
\ba
\chi &=&\frac{2\pi(\alpha_s-\alpha_{0})\sqrt{4\beta_{0}\beta_{2}-
\beta_{1}^{2}}}{2\pi(8\pi\beta_{0}+(\alpha_s+\alpha_{0})\beta_{1})+
\alpha_s\alpha_{0}\beta_{2}}. 
\ea

$\mathcal{ L}$ now corresponds to the LO solution,  which is the same as in  (\ref{truncatedseries}) 
while $\mathcal{M}$ and $\mathcal{Q}$ correspond to new functions of the running couplings at the two evolution 
scales, initial ($Q_0$) and final ($Q$). These two functions perform a resummation of the simpler logarithms included 
in (\ref{truncatedseries}). At NLO the implicit resummation performed by the exact non-singlet solution can be understood from its explicit form in $x$-space
\ba
f(\alpha_s,x)&=&\exp\left\{-\frac{2}{\beta_0}P^{(0)}(x)\log{\left(\frac{\alpha_s}{\alpha_0}\right)}
\right\}_\otimes \nonumber \\
&& \times\exp\left\{\left[\frac{2}{\beta_0}P^{(0)}(x) -\frac{4}{\beta_1}P^{(1)}(x)\right]
\log\left(\frac{4\pi\beta_0+\alpha_s \beta_1}{4\pi\beta_0+\alpha_0 \beta_1}\right)\right\}_\otimes f(\alpha_0,x)
\label{nlox1}
\ea
which generates the ordinary  contributions $\log{\left(\frac{\alpha_s}{\alpha_0}\right)}$ after an expansion
\ba
&&f(\alpha_s,x)=\exp\left\{-\frac{2}{\beta_0}P^{(0)}(x)\log{\left(\frac{\alpha_s}{\alpha_0}\right)}\right\}_\otimes
\times \nonumber\\
&&\hspace{1cm}\exp\left\{\left[\frac{2}{\beta_0}P^{(0)}(x)-\frac{4}{\beta_1}P^{(1)}(x)\right]
\left[\frac{\alpha_0\beta_1}{4\pi \beta_0+\alpha_0\beta_1}
\log{\left(\frac{\alpha_s}{\alpha_0}\right)}+\dots\right]
\right\}_\otimes f(\alpha_0). \nonumber \\
\ea

\section{LHC Phenomenology}
Both in the evolution performed with \textsc{Candia} and in the MRST and Alekhin evolution  \cite{MRST1,Alekhin} we use the same inputs given by these authors, choosing the initial scale $\mu_0^2=1.25$ GeV$^2$, and the same treatment of the heavy flavors, a rather nontrivial point from the numerical side. On the Z resonance we get for the NNLO  $K$-factors the values 
\ba
&&K(M_Z)=(\hat{\sigma}_{NNLO}\otimes\Phi^{NNLO}_{MRST})/(\hat{\sigma}_{NLO}\otimes\Phi^{NLO}_{MRST})=0.97\nonumber\\
&&K(M_Z)=(\hat{\sigma}_{NNLO}\otimes\Phi^{NNLO}_{\textsc{Candia}})/(\hat{\sigma}_{NLO}\otimes\Phi^{NLO}_{\textsc{Candia}})=0.95 \nonumber \\
&& K(M_Z)=(\hat{\sigma}_{NNLO}\otimes\Phi^{NNLO}_{Alekhin})/(\hat{\sigma}_{NLO}\otimes\Phi^{NLO}_{Alekhin})=0.98
\ea
where $\Phi$ denote the parton luminosities, which correspond to a reduction by $2.7 \%$ of the NNLO cross section compared to the NLO result (MRST evolution) and larger for the \textsc{Candia} evolution ($4.4 \% $), while for Alekhin is $1.5 \%$. From the analysis of the errors on the pdf's to NNLO, for instance for the Alekhin's set, the differences among these determinations 
render the results compatible, being the variations on the $K$-factors of the order of $4 \%$ 
(see the discussion in \cite{CCG2}).
\begin{figure}
\includegraphics[%
  width=6.5cm,
  angle=-90]{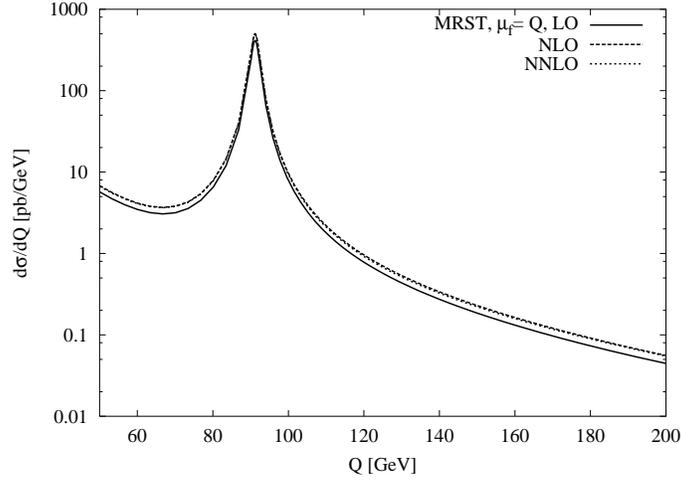}
\caption{Cross Sections in the region of the peak of the Z
boson at LO, NLO, and NNLO obtained using the
luminosities evolved respectively by MRST}
\label{Cross1}
\end{figure}

\begin{figure}
\includegraphics[%
  width=5.5cm,
  angle=-90]{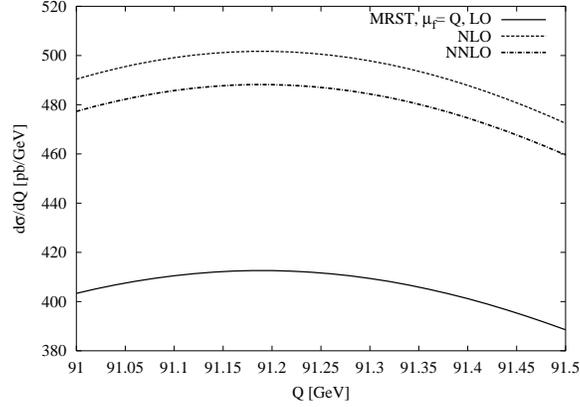}
\caption{Cross Sections in the region of the Z
with a zoom in the peak region.}
\label{Cross2}
\end{figure}
One can estimate
the role played by the evolution, respect to the NNLO hard scatterings by separating the two contributions 
in the variation of the hadronic cross section $\Delta \sigma$ between the NNLO and the NLO determinations, into contribution from the change 
of the hard scatterings  ($\Delta\hat{\sigma}$)  and from the luminosities ($\Delta \phi$)

\ba
\Delta\sigma=\Delta\hat{\sigma}\otimes\phi+\hat{\sigma}\otimes\Delta\phi. 
\ea
A numerical analysis shows that the change in the luminosities play a dominant role in 
the overall variation and are induced by the new terms in the NNLO evolution. 
Some of our results, taken from \cite{CCG2} are shown in Figs. \ref{Cross1} and \ref{Cross2}. The NNLO reduction for the invariant mass distributions is quite evident respect to the NLO result.
\section{Conclusions} 
In the cross sections that we have studied the NNLO QCD corrections are small (at the few percent level), but the reduction of the dependences from all the scales is significant. At the same time the issue of how we solve the evolutions equations is not secondary, given the small variations involved and the claimed accuracy. Given this issue and the problem with this systematic source of errors, our confidence on the parameterizations of the pdf's should never be too high, and  it is reasonable to keep a critical edge.  
We recall that with $10^{-1}$ fb of integrated luminosity the statistical error expected on the Z peak is
around $0.05 \%$ at the LHC, much smaller than the systematic error intrinsic to an NNLO prediction due to the evolution.
\centerline{\bf Acknowledgements} 
The work of C.C. was supported in part by INFN, the European Union through the Marie Curie Research and Training Network ``Universenet'' (MRTN-CT-2006-035863) and by the INTERREG IIIA Greece-Cyprus program, while A.C. is partly supported by 
the grant MTKD-CT-2004-014319 and by the EU grant MRTN-CT-2004-512194.  The work of M.G. is supported by INFN and MIUR.

%
%
%
%

\end{document}